\documentclass{aastex63}

\shorttitle{Important weak interaction nuclei}
\shortauthors{Nabi et al.}

\begin{document}

\title{INVESTIGATION OF IMPORTANT WEAK INTERACTION NUCLEI IN PRESUPERNOVA EVOLUTION}

\correspondingauthor{Jameel-Un Nabi}
\email{jameel@giki.edu.pk}
\email{vc@uow.edu.pk}

\author[0000-0002-8229-8757]{Jameel-Un Nabi}
\affiliation{GIK Institute of
	Engineering Sciences and Technology, Topi 23640, KP,
	Pakistan}
\affiliation{University of Wah, Quaid Avenue, Wah Cantt 47040, Punjab, Pakistan}

\author{Asim Ullah}
\affiliation{GIK Institute of
	Engineering Sciences and Technology, Topi 23640, KP,
	Pakistan}

\author{Ali Abas Khan}
\affiliation{FAST-National University of Computer and Emerging Sciences, Peshawar, KP, Pakistan}

\begin{abstract}

The project aims to investigate most important weak interaction nuclei in the presupernova evolution of massive stars. To achieve the goal, an ensemble containing 728 nuclei in the mass range A = 1--100 was considered. We computed the mass fractions of these nuclei using Saha's equation for predetermined values of \textit{$T$}, \textit{$\rho$} and \textit{$Y_e$} and assuming nuclear statistical equilibrium. The nuclear partition functions were obtained using a newly introduced recipe where excited states, up to 10 $MeV$, were treated as discrete. The weak interaction rates (electron capture (\textit{ec}) and $\beta$-decay (\textit{bd})) were calculated in a \textit{totally} microscopic fashion using the proton-neutron quasiparticle random phase approximation (pn-QRPA) model and without assuming the Brink-Axel hypothesis. The calculated rates were coupled with the computed mass fractions to investigate the  time rate of change of lepton to baryon fraction of the stellar matter. We compare our results with the previous calculations reported in the literature. Noticeable differences up to orders of magnitude are reported with previous calculations. These differences may influence the evolution of the star in the later stages of presupernova. We present a list of top 50 \textit{ec} and \textit{bd} nuclei which have the largest effect on $Y_e$ for conditions after silicon core burning. The competition between the \textit{ec} and \textit{bd} rates in stellar core was  investigated and it was found that $Y_e$ = 0.424--0.455 is the interval where the \textit{bd} results are bigger than the \textit{ec} rates. 
 
\end{abstract}


\section{Introduction} \label{sec:intro}
In the course of evolution of a massive star, weak interaction rates ($\beta$-decay ($bd$) and electron capture ($ec$)) play crucial roles. These reactions contribute to the gravitational collapse of the core of a massive star by reducing the electron degenerate pressure, thereby causing a supernova explosion and thus affect the formation of massive (neutron-rich) nuclei \citep{bethe1979equation, bethe1990supernova}.   

Several attempts have been made for the calculation of weak interaction rates in stellar environment in a bid to have a better understanding of the stellar evolution. Fuller, Fowler, and Newman (FFN) \citep{fuller1980stellar, fuller1982stellar, fuller1985stellar} employed the independent-particle model (IPM) using the then available experimental data and tabulated the weak interaction rates for 226 nuclei with mass in the range 60 $\ge$ A $\ge$ 21. Their rates led to a considerable reduction in the lepton to baryon fraction ($Y_e$) throughout the stellar core and  significantly contributed to a better understanding of the presupernova evolution of stars \citep{weaver1985numerical}. In 1994, Aufderheide and collaborators \citep{aufderheide1994search} studied the influence of weak interaction in evolution of massive stars post silicon burning and searched for most important \textit{ec} and \textit{bd} nuclei in $Y_e$ range of  0.400--0.500 using the IPM model. \textbf{They extended the FFN work for heavier nuclei
with A $>$ 60 and took into consideration explicit quenching
of the GT strength}. In the same decade, many authors e.g. \citep{vetterli1989gamow, el1994spin, williams1995gamow} raised concern on the systematic parameterization employed by FFN and later adopted by Aufderheide et al. 
\textbf{The proton-neutron quasiparticle random phase approximation (pn-QRPA) \citep{nabi1999weak, nabi2004microscopic} and shell model \citep{langanke2000shell} later calculated weak interaction rates in stellar matter and showed that primarily the misplacement of the GT centroids by FFN in some key nuclei led to disagreement with experimental values.} 
Heger and collaborators \citep{heger2001presupernova} utilized weak rates in the mass range A $ = 45-65$ based on large-scale shell-model (LSSM) \citep{langanke2001rate} and performed simulation studies during the presupernova evolution. A similar study of the presupernova evolution using the pn-QRPA rates was not performed and was much awaited.

The aim of the current work is to search for the most important weak interaction nuclei in the presupernova evolution of massive stars using the pn-QRPA model. To achieve this goal, we considered a large pool containing 728 nuclei having atomic number up to 50 and mass number up to 100, in order to cover a decent number of nuclei prevailing in the stellar matter. The nuclei that contribute to largest change in $Y_e$ values are neither the most abundant  nor the ones with the strongest rates but a combination of the two. Thus one must know both the rates and nuclear abundance of nuclei to determine the most important weak interaction (\textit{ec} or \textit{bd}) nuclei for given values of  \textit{$T$} (temperature),  \textit{$\rho$} (baryon density) and \textit{$Y_e$}. To compute \textit{ec} and \textit{bd} rates for the wide suite of selected nuclei, we employed the  pn-QRPA~\citep{halbleib1967gamow} theory, a
successful microscopic model frequently used in the past, with reasonable success, to compute the weak rates
under terrestrial \citep{staudt1990second, hirsch1993microscopic} and stellar conditions \citep{nabi1999weak, nabi2004microscopic}. The reliability of the pn-QRPA model was thoroughly discussed earlier~\citep{nabi2004microscopic}. A very decent comparison of pn-QRPA model calculation with experimental data was achieved for more than thousand nuclei (see Tables~E--M and Figures~11--17 of \citep{nabi2004microscopic}. For further comparison of pn-QRPA model with measured data please see Refs.~\citep{majid2016study, majid2017allowed, nabi2016beta, nabi2015first, nabi2020nuclear}). The pn-QRPA model is much suitable for stellar rate calculations for the important reason that it does not assume the Brink-Axel hypothesis \citep{brink1958theory} for computation of excited state GT strength distributions, as usually done in such calculations. Furthermore a large model space,  up to seven major shells, makes the model calculation possible for any arbitrary heavy nucleus. Nuclear abundances were determined using the Saha's equation assuming nuclear statistical equilibrium (NSE). We later investigate  the total rate of change of $Y_e$ with changing stellar conditions and also compare our results with previous calculations.

This paper is structured as follows: In Section~2 we briefly introduce the
formalism employed to compute weak rates and  nuclear abundances. Our results are
discussed and compared with previous calculations
in Section~3. Finally, concluding remarks are stated in the last section.

\section{Formalism} \label{sec:formalism}
\subsection{Weak interaction rates} \label{subsec:rates}

The \textit{ec} and \textit{bd} rates from the $\mathit{i}$th state of the parent to
the $\mathit{j}$th state of the daughter nucleus in stellar matter is given by

\begin{equation}\label{rate}
\lambda_{ij} =ln2
\frac{\phi_{ij}^{ec(bd)}(T,\rho,E_{f})}{(ft)_{ij}}.
\end{equation}

In Eq.~\ref{rate}, $(ft)_{ij}$ is related to the reduced transition
probability $B_{ij}$ of the nuclear transition by

\begin{equation}\label{ft}
(ft)_{ij}^{ec(bd)}=D/B_{ij},
\end{equation}
where 
\begin{equation}
B_{ij}=B(F)_{ij}+(g_{A}/g_{V})^2 B(GT)_{ij}.
\end{equation}
The value of constant D was taken as
6143 \textit{s} \citep{hardy2009superallowed} and the ratio of axial-vector and vector coupling constant was kept as $g_{A}$/$g_{v}$= -1.2694 \citep{nakamura2010review}. $B(F)$ and $B(GT)$ are the reduced transition probabilities of the Fermi and GT transitions, respectively.
\begin{equation}
B(F)_{ij} = \frac{1}{2J_{i}+1} \mid<j \parallel \sum_{k}t_{\pm}^{k}
\parallel i> \mid ^{2}
\end{equation}

\begin{equation}\label{bgt}
B(GT)_{ij} = \frac{1}{2J_{i}+1} \mid <j
\parallel \sum_{k}t_{\pm}^{k}\vec{\sigma}^{k} \parallel i> \mid ^{2},
\end{equation}
where $J_i$ is the spin of  parent state,
$\vec{\sigma}$ are the Pauli spin matrices and $t_\pm$ refer
to the isospin raising and lowering operators.

The phase space function $(\phi_{ij})$ is an integral over the total
energy. For the case of \textit{bd}, it is given by (here
onwards we use natural units, $\hbar=m_{e}=c=1$)
\begin{equation}\label{ps}
\phi_{ij} = \int_{1}^{\omega_{m}} \omega \sqrt{\omega^{2}-1} (\omega_{m}-\omega)^{2} F(+ Z,\omega)
(1-G_{-}) d\omega,
\end{equation}
whereas  the phase space for \textit{ec} is given by
\begin{equation}\label{pc}
\phi_{ij} = \int_{\omega_{l}}^{\infty} \omega \sqrt{\omega^{2}-1} (\omega_{m}+\omega)^{2} F(+
Z,\omega) G_{-} d\omega,
\end{equation}
where  $\omega$ is the
total energy of the electron including its rest mass, $\omega_l$ is the total capture threshold energy for \textit{ec} and $ \omega_{m}$ is the total \textit{bd} energy. $\omega_{m}  = (m_{p}-m_{d}+E_{i}-E_{j}$) where $m_p (m_d)$ and $E_i (E_j)$ are the mass and excitation energies of
the parent (daughter) nucleus.  The  $F(+ Z,\omega)$ are the Fermi
functions computed using the recipe of Gove and Martin ~\citep{gove1971log}. $G_{-}$
is the electron distribution function given by
\begin{equation}
G_- = [exp(\frac{E-E_f}{kT})+1]^{-1},
\end{equation}
where $E = (\omega-1)$ stands for kinetic energy of electrons and $E_f$ is the Fermi energy of electrons.

The number density of protons associated with electrons and nuclei
was determined using
\begin{equation}\label{ye}
\rho Y_{e} = \frac{1}{\pi^{2}N_{A}}(\frac {m_{e}c}{\hbar})^{3}
\int_{0}^{\infty} (G_{-}-G_{+}) p^{2}dp,
\end{equation}
where $p=(\omega^{2}-1)^{1/2}$ is lepton's momentum and $G_{+}$
is the positron distribution function
\begin{equation}
G_+ = [exp(\frac{E+2+E_f}{kT})+1]^{-1}.
\end{equation}

Due to prevailing high temperatures in stellar core, there is a finite probability of occupation of parent excited states. Hence weak decays have finite contribution from these states. Assuming thermal equilibrium, the probability of occupation of parent state $i$ was estimated using 

\begin{equation}\label{pi}
P_{i} = \frac {exp(-E_{i}/kT)}{\sum_{i=1}exp(-E_{i}/kT)}.
\end{equation}
The total weak-decay rate per unit time per nucleus
was finally computed using
\begin{equation}\label{ecbd}
\lambda^{ec(bd)} =\sum _{ij}P_{i} \lambda^{ec(bd)} _{ij}.
\label{total rate}
\end{equation}
\textbf{We considered 300 parent ad 300 daughter excited states in our rate calculation.} The summation over the initial and final states was carried out
until satisfactory convergence was achieved in the
calculation.  $P_{i} $ is the probability of occupation of
parent excited states and follows the normal Boltzmann distribution.

\subsection{Nuclear abundances} \label{subsec:NSE}
During stellar evolution, once the silicon burning is completed, NSE can be achieved where strong and electromagnetic interactions are in equilibrium.  The weak interaction  is still not in equilibrium as the stellar matter is transparent to neutrinos produced and energy is channeled out of the system. In such a scenario, the isotropic abundances of nuclei follow simply from the nuclear Saha's equation for a given $\rho$, $T$ and $Y_{e}$ \citep{hartmann1985nucleosynthesis}. We treated NSE in the same way it was previously employed by \citep{ clifford1965equilibrium, kodama1975r, hartmann1985nucleosynthesis}.

According to Saha's equation, the nuclear abundance of k\textit{th} nucleus is given by 
\begin{equation}\label{x1}
\tau_k(A,Z)= \frac {C(A,Z,T)}{2}\left(\frac{\rho N_{A} \lambda_T^{3}}{2}\right)^{A-1}\times A^{\frac{5}{2}}{\tau_{n}^{A-Z}\tau_{p}^{Z}}\exp
\left[ Q_k(A,Z)/k_BT \right],
\end{equation}
where $C$(A,Z,T) is the  nuclear partition function of the k\textit{th} nucleus, $N_A$ is the Avogadro's number, $\lambda_T$ (=$\sqrt{\frac{h^2}{2\pi m_Hk_BT}}$) is the thermal wavelength, $Q_k$ is the ground state binding energy and $k_B$ is the Boltzmann constant.  $\tau_n$ and $\tau_p$ are the mass fractions of the free neutrons and protons, respectively,  and can be determined subject to mass conservation 
\begin{equation}
\sum_{k} \tau_k = 1
\end{equation} 
and charge conservation
\begin{equation}
\sum_{k} \frac{Z_k}{A_k}\tau_k = Y_e = \frac{1-\eta}{2},
\end{equation}
where $\eta$ is the neuron excess.  $\tau_k$ for k\textit{th} nucleus can be calculated once $\tau_n$ and $\tau_p$ are determined. The time derivative of $Y_e$ is a key parameter to be monitored during the presupernova evolution and for a \textit{k}th nucleus it is given by
\begin{equation}
\dot{Y}_{e(k)}^{ec(bd)} = -(+)\frac{\tau_k}{A_k}\lambda_k^{ec(bd)},
\end{equation}
where the negative sign is for \textit{ec} and positive for \textit{bd}, $A$ is the mass number and $\lambda^{ec(bd)}$ was calculated using  Eq.~\ref{ecbd}.

While investigating isotopic abundances in presupernova cores \citep{dimarco2001influence} showed that the computed nuclear partition functions showed a deviation
of up to 50$\%$ when low-lying nuclear states were considered as discrete energy levels as against those assuming a level density function and performing
integrations. The authors were able to conduct a summation over states up to 3 $MeV$. Since the computed nuclear partition function is the cause of one of the biggest uncertainties in calculated mass fractions, Nabi and collaborators \citep{aliMF,nabi2016mass} recently introduced a novel recipe for an accurate description of nuclear partition functions which we follow here. The basic idea is to calculate discrete energy levels up to 10 $MeV$ excitation energies. Beyond 10 $MeV$, a simple level
density function was assumed and integration was performed
up to 25 $MeV$ excitation energy. In case few measured levels were missed in the process, they were inserted manually along with their spins in
the calculation.

The nuclear partition function was calculated using 
\begin{equation}\label{ff}
C(A,Z,T)=\sum_{\mu = 0}^{\mu_{m}}(2J^{\mu}+1)exp[{-E^{\mu}/kT}]+\int_{E^{\mu_{m}}}^{E^{max}}\sum_{J^{\mu}, \pi^{\mu}}(2J^{\mu}+1)exp(-\epsilon/kT)\rho_l(\epsilon, J^{\mu},\pi^{\mu})d\epsilon,
\end{equation}  
where $\mu_{m}$ labels the last employed (either experimentally available or theoretically calculated) energy eigenstate. The first term in the above equation represents contribution from  low-lying states (experimental or pn-QRPA predicted energy eigenvalues) up to $E^{\mu}$ and the sum runs over all Boltzmann-weighted states from ground state to $\mu_m$. The second term uses process of integration to sum the contribution from continuous states. $E^{max}$ was taken as 25 $MeV$. For details of the formalism we refer to \citep{aliMF,nabi2016mass}.

\section{Results and Discussion} \label{sec:RandD}
\begin{figure}[ht!]
	\plotone{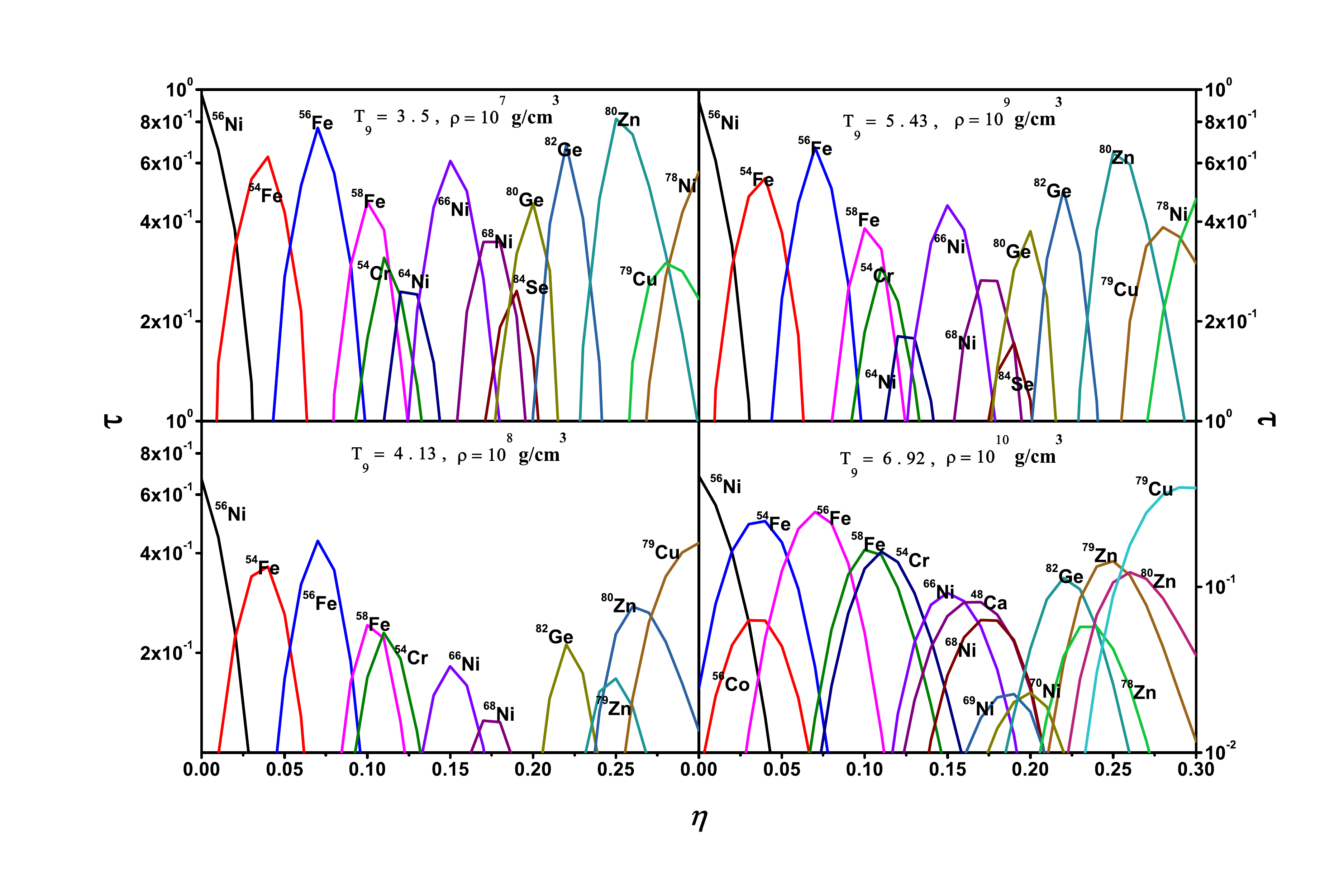}
	\caption{Mass fractions of some abundant nuclei in NSE as a function of neutron excess, at different densities and temperatures. \label{mf} }
\end{figure}

\begin{figure*}
	\plotone{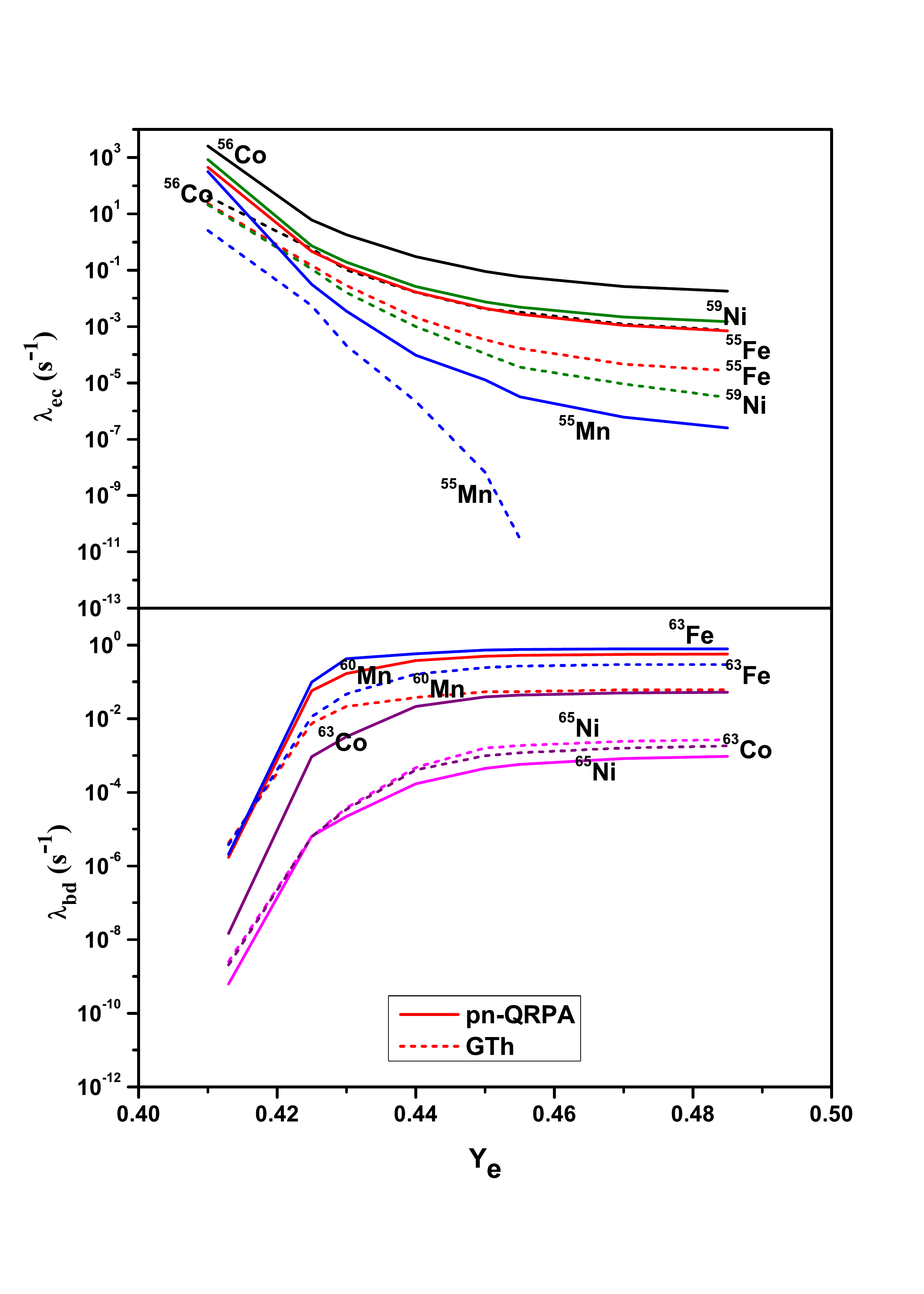}
	\caption{Comparison of the pn-QRPA and GTh  weak decay rates as a function of $Y_e$.
		\label{RateBDEC}}
\end{figure*}

\begin{figure}[ht!]
	\plotone{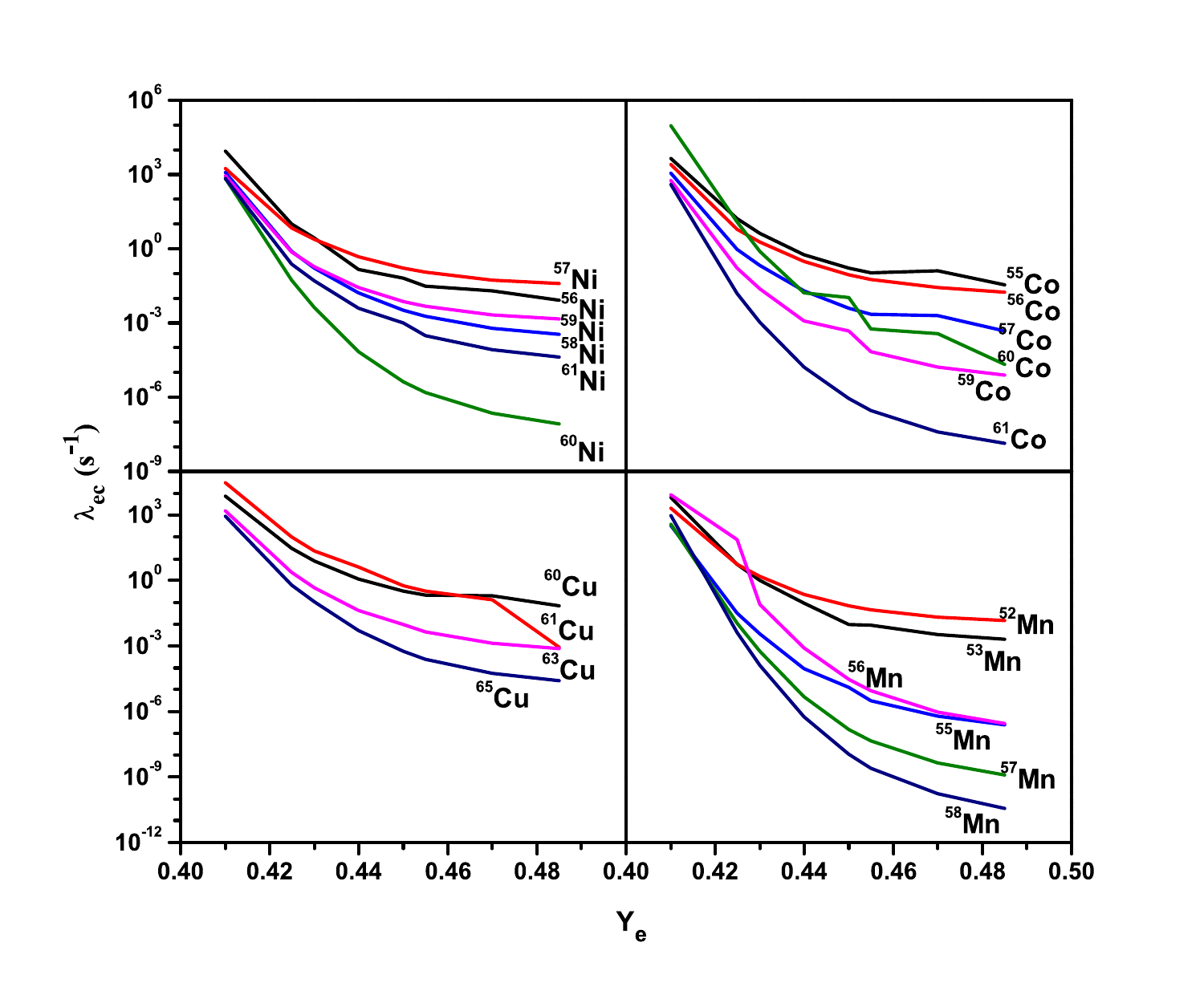}
	\caption{The pn-QRPA calculated \textit{ec} rates as a function of $Y_e$ for isotopes of Ni, Co, Cu and Mn. \label{RateYeEC}}
\end{figure}

\begin{figure}[ht!]
	\plotone{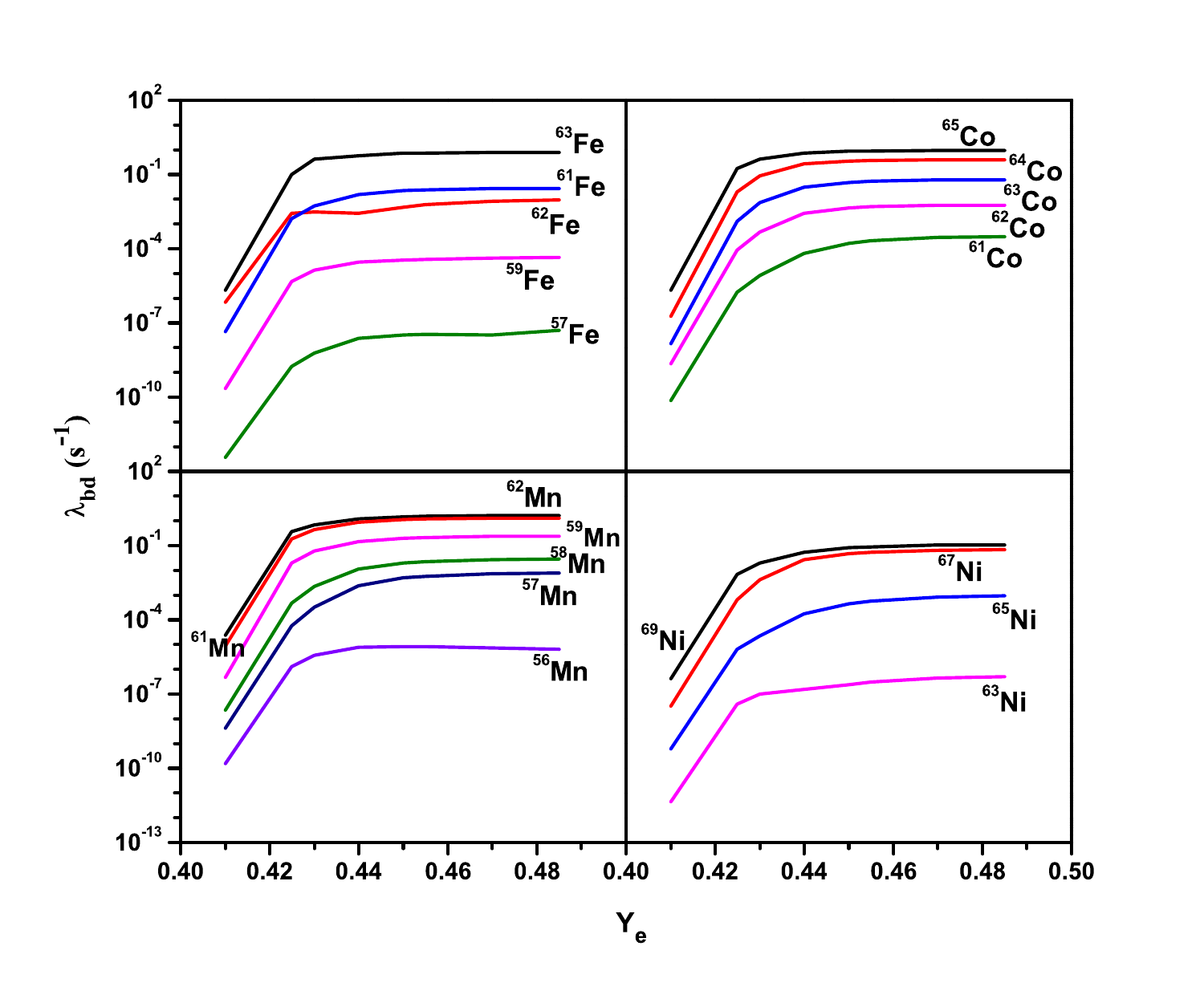}
	\caption{The pn-QRPA calculated \textit{bd} rates as a function of $Y_e$ for isotopes of Fe, Co, Mn and Ni. \label{RateYeBD}}
\end{figure}

\begin{figure*}
	\plotone{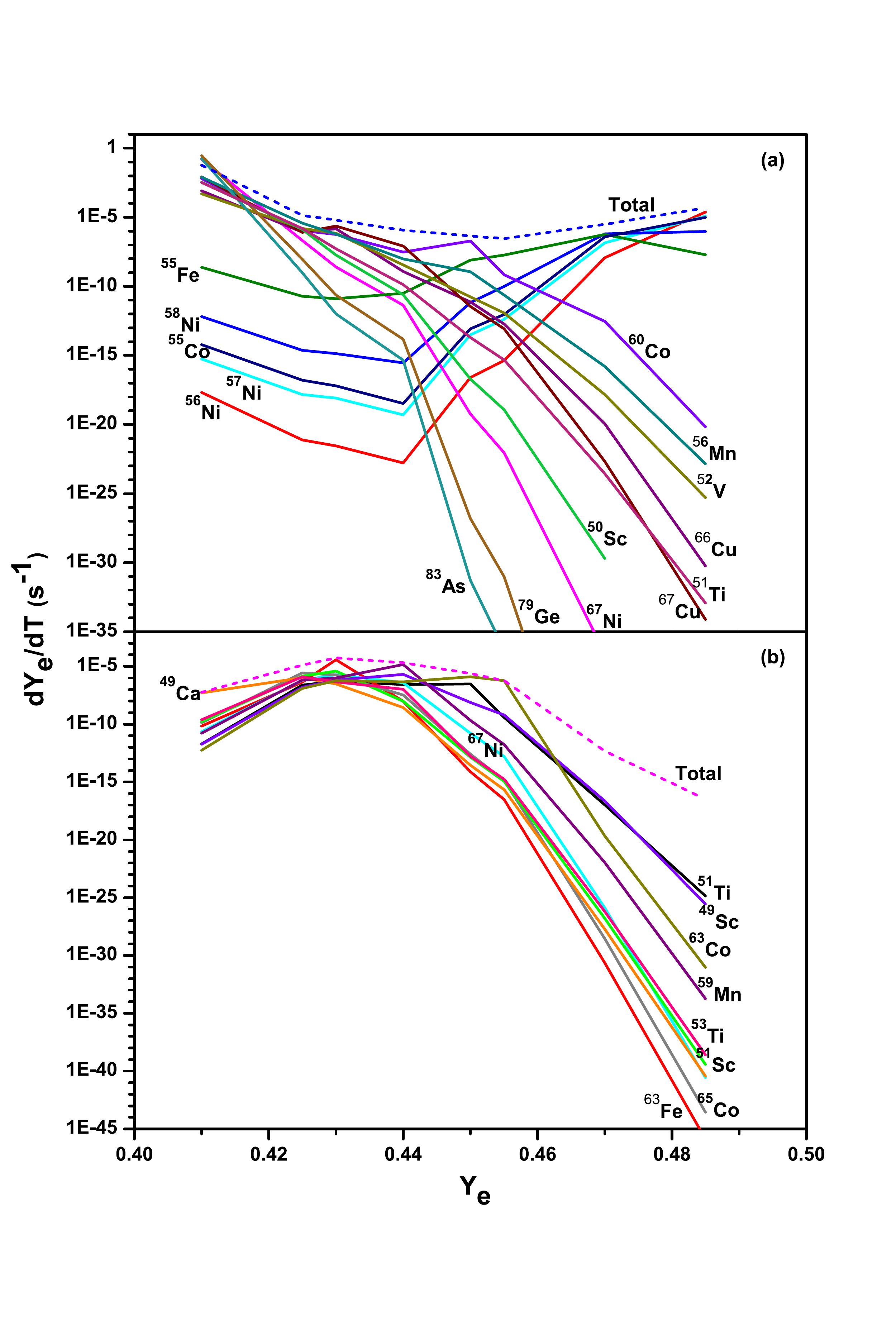}
	\caption{Evolution of temporal derivative of $Y_e$ for (a) \textit{ec} and (b) \textit{bd} rates.
		\label{YdYeBDCD}}
\end{figure*}

\begin{figure}[ht!]
	\plotone{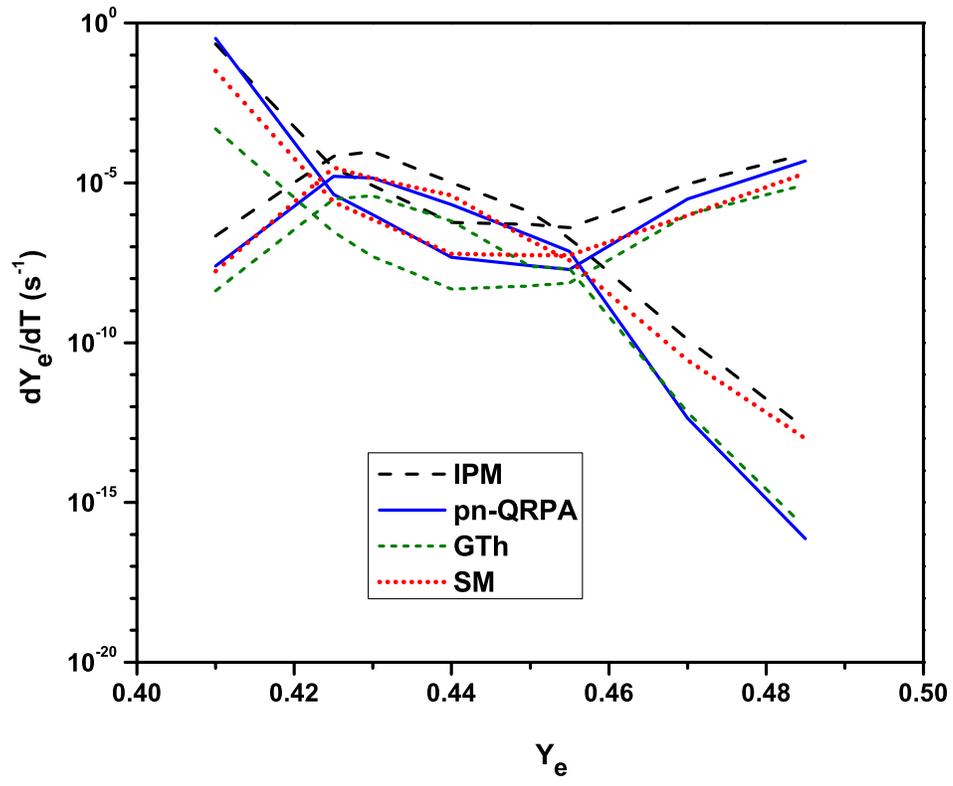}
	\caption{Competition between \textit{ec} and \textit{bd} in the evolution of $\dot{Y}_e$. \label{ECvBD}}
\end{figure}

We employed the pn-QRPA model to compute the weak decay rates  and mass fractions for our selected pool containing  728 nuclei in the mass range A = 1--100.  It is worth mentioning that our  pool of nuclei is significantly larger than the ones considered by previous studies. Recently the Gross Theory \citep{ferreira2014weak} was employed to estimate \textit{ec} and \textit{bd} rates under stellar conditions. We compared our calculation with the results of Gross Theory (referred to as GTh throughout the text) and the IPM calculation performed by  \citep{aufderheide1994search} and referred to as IPM onwards. Our selected pool of 728 nuclei may be compared with 150, 63 and around 100 nuclei chosen by IPM, GTh and SM~\citep{langanke2000shell}, respectively. A small pool of nuclei can omit few key nuclei which may
dominate the nuclear composition during the late phases of the collapse before neutrino trapping.  The number of nuclei in stellar ensemble may reach around 2700 species, as reported by the authors in Ref. \citep{juodagalvis2010improved}. According to previous findings \citep{rauscher2000astrophysical, rauscher2003nuclear}, calculation of partition functions is sensitive to the input mass models. All theoretical masses used in this project (including Q-values)  were derived from the recent mass compilation of Audi et al. \citep{audi2017nubase2016}. For the cases where mass values were not given, we used Ref. \citep{moller2016nuclear} to compute the Q-values. \\
We estimated the rate of change of lepton to baryon fraction ($\dot{Y}_e$), separately for \textit{ec}  (Tables \ref{T1}-\ref{T3}) and \textit{bd} rates (Tables \ref{T4}-\ref{T6}) and sorted 30 nuclei with largest contribution to $\dot{Y}_e$. The \textit{ec} and \textit{bd} rates, for all the nuclei listed in Tables \ref{T1}-\ref{T6},  have been calculated using only allowed GT transitions. 
Our calculated weak rates and the corresponding $\dot{Y}_e$ values are, most of the times smaller than, sometimes bigger and at times comparable to the corresponding results of IPM. For example, our calculated \textit{ec} rates are smaller than  IPM rates by a factor of 4  for $^{50}$V, $^{54}$Fe and $^{64}$Cu and factor 2 for $^{61}$Co, $^{58}$Co, $^{59}$Ni and $^{63}$Ni (see Table \ref{T1}-\ref{T3}). Our \textit{bd} rates are a factor 3, 4 and 5 bigger than IPM rates for $^{53}$Ti, $^{66}$Co and $^{65}$Co, respectively (Table \ref{T4}-\ref{T6}). The biggest difference between the two calculations were noted for $^{54}$Mn and $^{62}$Cu ($^{58,60}$Co), where our computed \textit{ec} (\textit{bd}) rates were found 2 (5) orders of magnitude smaller than IPM results. Similar differences were found for $\dot{Y}_e$ values between the two models. Our computed total $\dot{Y}_e$ (shown in the bottom of Tables~\ref{T1}-\ref{T6}) is always smaller than IPM value both at low and high ($T_9$, $\rho$) values. \textbf{Misplacement of GT centroids} relative to experimental values led to bigger weak rates in majority of cases for IPM.  When comparing our \textit{ec} rates with those calculated by GTh, it is noted that \textit{all} rates and the corresponding $\dot{Y}_e$ values calculated by our model are bigger by orders of magnitude. Only in very few \textit{bd} cases  ($^{49}$Ca, $^{56}$Mn and $^{60}$Co), our computed rates and the respective $\dot{Y}_e$ values are smaller than GTh.  The GTh results include transitions \textit{only} from ground state which clearly is a poor approximation to be used for stellar rate calculations. The pn-QRPA model included a huge model space of 7$\hbar \omega$ to efficiently handle excited states in parent and daughter nuclei. Keeping in view the good comparison of pn-QRPA calculated values with measured data, our  ground and excited state GT strength distributions depict a  more realistic picture of the proceedings and contribute significantly to the reliability of our calculation. We again remind that Brink-Axel hypothesis was not used in computation of excited state GT strength functions in our calculation.    \\ 
Figure \ref{mf} shows our computed mass fractions for few  abundant nuclei in NSE as a function of neutron excess ($\eta$) at different temperatures and densities. For the sake of comparison we used the same parameters ($T_9$, $\rho$ and \textit{$\eta$}) as used in Figure 1 of IPM. It is noted that the computed mass fractions follow the same behavior in both models with slight variations. The different recipes for nuclear partition functions and
different mass models in the two approaches resulted in this small difference. We believe that using our recipe of nuclear partition functions would lead to a more reliable computation of mass fractions (see also \citep{nabi2016mass} for detailed analysis  of the comparison between the two computed mass fractions).\\
Figure \ref{RateBDEC} compares our computed \textit{ec}  (upper panel) and \textit{bd} (lower panel) rates as a function of $Y_e$ with the GTh results. The \textit{ec} rates are  sensitive to the stellar temperature and density values, showing an increase for higher values of $T_9$ and $\rho$. The behavior of our computed weak rates as a function of $Y_e$ is also in  agreement with IPM  i.e. the \textit{ec} rates approaching higher values when the magnitude of $Y_e$ is reduced ($T_9$ and $\rho$ increased) while there is a decrease in \textit{bd} rates when the $Y_e$ is decremented. The electron chemical potential increases with rise in stellar density and that is why the \textit{ec} rates are enhanced at lower $Y_e$ values. For higher density, the enhanced chemical potential of electron outside the nuclei impede the \textit{bd} rates. The major reduction in the density (4E+10 to 1E+9)  $g/cm^3$ accounts for the sudden increase (decrease) in the \textit{bd} (\textit{ec}) rates in the short interval of $Y_e$ = 0.410-0.430. The slope of the graph is relatively less steep for higher $Y_e$ values because of the smooth reduction in the $T_9$ and $\rho$ values. Our computed weak rates are bigger by as much as 1--2 orders of magnitude when compared with those of GTh results for the cases shown in Figure~\ref{RateBDEC}. The reasons for our bigger rates were discussed above. Figure \ref{RateYeEC} depicts our computed \textit{ec} rates, as a function of $Y_e$, for Ni, Co, Cu and Mn isotopes while Figure \ref{RateYeBD} shows our \textit{bd} rates for  Fe, Co, Ni and Mn isotopes. \\
The evolution of rate of change of lepton to baryon fraction ($\dot{Y}_e$) is shown in Figure \ref{YdYeBDCD}. Panel (a) depicts 15 nuclei with largest contribution to the total $\dot{Y}_e$ values for \textit{ec}, whereas panel (b) shows similar results for 10 nuclei having higher $\dot{Y}_e$ values for \textit{bd}. Figure \ref{ECvBD} shows the sum of $\dot{Y}_e$ values for both \textit{ec} and \textit{bd} rates. Here we compare our results with three other models namely IPM, GTh and SM.   It can be seen from Figures \ref{YdYeBDCD}-\ref{ECvBD} that the \textit{ec} rates dominate toward the end values of our chosen range of $Y_e$, while the \textit{bd} rates win somewhere at the center value ($Y_e$ $\approx$ 0.440). At $Y_e$ = 0.410, the contribution of $^{67}$Ni, $^{83}$As and $^{79}$Ge to the total $\dot{Y}^{ec}_{e}$ is greater than the contribution of $^{49}$Ca, $^{63}$Fe and $^{51}$Sc to the $\dot{Y}^{bd}_{e}$ and hence the \textit{ec} rates are dominant at this point. The $\dot{Y}^{ec}_{e}$ decreases and $\dot{Y}^{bd}_{e}$ increases with rise in $Y_e$ values up to $Y_e$ $\approx$ 0.422. However the \textit{ec} rates are still dominant. With further increase in the $Y_e$ value beyond 0.422, though the contribution of $\dot{Y}^{ec}_{e}$
for $^{67}$Cu, $^{56}$Mn and $^{51}$Ti is bigger, once the share of other key \textit{bd} nuclei are summed up, the total $\dot{Y}^{ec}_{e}$ is smaller than the corresponding $\dot{Y}^{bd}_{e}$. The increase (decrease) in $\dot{Y}^{bd}_{e}$ ($\dot{Y}^{ec}_{e}$) continues till $Y_e$ = 0.440 (0.455). Beyond $Y_e$ $\approx$ 0.440, where the density is 3.30E+8 $g/cm^3$ and temperature is around 4.24E+9 $K$, the $\dot{Y}^{bd}_{e}$ starts to decrease but still dominates up to values of $Y_e$ $\approx$ 0.455 due to main contribution coming from $^{63}$Co and $^{51}$Ti. The decrease in \textit{bd} continues till end point ($Y_e$ $\approx$ 0.500) whereas the \textit{ec} continues growing beyond $Y_e$ $\approx$ 0.455 with major contribution coming from $^{55}$Fe, $^{56-58}$Ni and $^{55}$Co. The \textit{bd} remains dominant over \textit{ec} for the interval $Y_e$ $\approx$ 0.424-0.455, 11\% greater than the one proposed by IPM ($Y_e$ $\approx$ 0.425-0.453). Our proposed interval was found close to the SM and GTh intervals, which were $Y_e$ $\approx$ 0.42-0.46 and $Y_e$ $\approx$ 0.422-0.455,  respectively. It may be seen from Fig. \ref{ECvBD} that our estimated total $\dot{Y}^{ec(bd)}_{e}$ values are smaller in magnitude than IPM results, bigger than GTh estimations and in good agreement with the SM estimation. The end result  necessitates the use of a microscopic model for calculation of stellar weak rates. \\ 
We finally determined the most important weak interaction nuclei that have the largest effect on ${Y}_{e}$ post silicon core burning,  by averaging the contribution from each nucleus to $\dot{Y}_{e}$ over the entire chosen stellar trajectory. To this effect we introduced a ranking parameter ($\mathring{R}_p$) given by
\begin{equation}
 \mathring{R}_{p} = \left(\frac{\dot{Y}^{ec(bd)}_{e(k)}}{\sum\dot{Y}^{ec(bd)}_{e(k)}}\right)_{0.500 > {Y_e} > 0.400}	
\end{equation}
Thus, nuclei with largest contribution to $\dot{Y_e}$ will have the highest value of $\mathring{R}_p$. On the basis of $\mathring{R}_p$, we listed top 50  \textit{ec} and \textit{bd} nuclei for all conditions which follow silicon core burning in Table \ref{T7}. 
It is worth noting that 90\% (70\%) of our most important \textit{ec} (\textit{bd}) nuclei are the same as the list provided by IPM.  The nuclei marked with an asterisk are the ones missing in the list provided by the IPM.

\section{Concluding Remarks} \label{sec:C}

In this work we present the first ever compilation of nuclei which have the largest effect on ${Y}_e$ for conditions after silicon core burning using the pn-QRPA model. The use of a fully microscopic model adds to the reliability of calculated weak rates. The pn-QRPA model does not employ the Brink-Axel hypothesis for calculation of excited state GT strength distributions -- a feature not available even in the large scale shell model calculation. We are currently working on quantifying the disadvantages of employing Brink-Axel hypothesis for calculation of stellar rates and hope to report our findings soon. The reliability of pn-QRPA results and its decent comparison with measured data were discussed in Section~1. Our computation of mass abundances also used a novel recipe for calculation of nuclear partition functions. We considered all excited states up to 10 $MeV$ as discrete which may result in a more realistic calculation of isotopic abundances according to a previous study.  

Our results are shown in Tables~1-7 which show  rankings of nuclei during several key snapshots of presupernova evolution of massive stars. Specially Table~7 presents an overall ranking for all conditions post silicon burning phase and we compile a list of top 50  \textit{ec} and \textit{bd} nuclei causing largest effect on ${Y}_e$. It is to be noted that the list contains 10\% of new \textit{ec} and 30\% of new \textit{bd} nuclei not to be seen in the list of IPM tables. \textbf{For a nucleus with A $\sim$ 100 at $T_9$ = 5.39 \textit{GK}, the mean nuclear excitation energy is $\sim$ 3 \textit{MeV} and transitions above 10 \textit{MeV} excitation is in order. Our nuclear partition function (Eq. 17) considers discrete states only up to 10 \textit{MeV}. Beyond 10 \textit{MeV} we assumed a level density function and this may bear consequences for the calculated mass fractions.} Core-collapse simulators may find this updated list interesting for their future studies. This list can further be extended to top 700 nuclei and may be requested from the corresponding author.

Another key finding of this paper is that we propose an updated interval $Y_e$ = 0.424-0.455 in which the \textit{bd} rates dominate the competing \textit{ec} rates which is roughly 11\% bigger than proposed by IPM and 6\% smaller than GTh range. Further our estimated total $\dot{Y}^{ec(bd)}_{e}$ values are in good agreement with shell model prediction. We plan to extend our pool of nuclei with $A >$ 100 and explore role of forbidden transitions (that become important for neutron-rich matter) in the list of nuclei having largest effect on $Y_e$ in the near future.

\section{Acknowledgments} \label{sec:d}
J.-U. Nabi would like to acknowledge the support of the Higher
Education Commission Pakistan
through project numbers 5557/KPK
/NRPU/R$\&$D/HEC/2016 and 9-5(Ph-1-MG-7)/PAK-TURK
/R$\&$D/HEC/2017.

\begin{deluxetable}{cc|ccc|ccc}
	\tablenum{1}
	\tablecaption{The \textit{ec} rates for most important nuclei sorted in order of $\mid$$\dot{Y}^{ec}_e$$\mid$. The rates are compared with IPM and GTh results whereever available.  The units of $\rho$, $T_9$, $\lambda^{ec}$ and $\dot{Y}^{ec}_e$ are $g/cm^3$, \textit{GK}, $s^{-1}$ and $s^{-1}$, respectively.  }
	\tablewidth{0pt}
	\tablehead{
		\multicolumn1c{} & \multicolumn2c{$\rho$ = 1.06E+09,} &\multicolumn2c{$T_9$ = 4.93,}  &
		\multicolumn{2}{c}{$Y_e$ = 0.430} & \multicolumn1c{}  \\
		\hline
		\colhead{} & \colhead{} &\multicolumn3c{ {$\lambda_k^{ec}$}}  &
		\multicolumn{3}{c}{$\mid$$\dot{Y}^{ec}_{e(k)}$$\mid$}  \\
		\colhead{A} & \colhead{Symbol} & pn-QRPA & IPM & GTh &
		\colhead{pn-QRPA} & \colhead{IPM} & \colhead{GTh}
	}
	\startdata
	67 & Cu & 3.54E-02 & 2.38E-03 & ---      & 2.33E-06 & 2.26E-07 & ---      \\
	66 & Cu & 4.63E-01 & 6.56E-01 & ---      & 1.47E-06 & 1.30E-06 & ---      \\
	52 & V  & 5.64E-02 & 1.81E-02 & 1.17E-05 & 7.05E-07 & 2.58E-07 & 1.16E-08 \\
	56 & Mn & 7.91E-02 & 2.90E-02 & 2.61E-04 & 5.88E-07 & 2.91E-07 & 1.28E-08 \\
	60 & Co & 7.69E-01 & 2.31E+00 & 4.27E-03 & 5.65E-07 & 3.66E-06 & 1.80E-08 \\
	49 & Sc & 1.94E-03 & ---     & ---      & 2.63E-07 & ---      & ---      \\
	65 & Cu & 1.08E-01 & 5.91E-02 & ---      & 1.43E-07 & 1.67E-07 & ---      \\
	62 & Co & 9.31E-03 & 3.22E-02 & 2.75E-05 & 1.26E-07 & 5.34E-07 & 1.69E-09 \\
	54 & Cr & 6.41E-05 & ---      & ---      & 7.10E-08 & ---      & ---      \\
	49 & Ti & 6.41E-02 & 1.70E-02 & ---      & 5.67E-08 & 4.09E-08 & ---      \\
	53 & Cr & 5.16E-03 & ---      & ---      & 5.13E-08 & ---      & ---      \\
	51 & Ti & 1.08E-04 & ---      & ---      & 5.13E-08 & ---      & ---      \\
	51 & V  & 6.34E-03 & 9.71E-03 & 1.67E-04 & 4.87E-08 & 1.53E-07 & 2.32E-09 \\
	63 & Ni & 1.92E-03 & 3.87E-03 & ---      & 4.51E-08 & 8.15E-08 & ---      \\
	64 & Ni & 2.10E-05 & ---      & ---      & 4.25E-08 & ---      & ---      \\
	48 & Sc & 8.34E-02 & 8.25E-02 & ---      & 4.05E-08 & 1.02E-07 & ---      \\
	57 & Mn & 5.78E-04 & 4.03E-04 & ---      & 3.83E-08 & 6.33E-08 & ---      \\
	61 & Co & 1.09E-03 & 2.23E-03 & ---      & 3.79E-08 & 2.42E-07 & ---      \\
	68 & Cu & 2.18E-03 & 2.62E-02 & ---      & 3.66E-08 & 3.21E-07 & ---      \\
	55 & Cr & 1.66E-04 & ---      & ---      & 2.51E-08 & ---      & ---      \\
	73 & Ga & 4.58E-03 & ---      & ---      & 2.26E-08 & ---      & ---      \\
	59 & Fe & 1.36E-04 & 1.83E-04 & ---      & 1.92E-08 & 4.51E-08 & ---      \\
	53 & V  & 2.74E-04 & ---      & ---      & 1.90E-08 & ---      & ---      \\
	50 & Sc & 1.03E-03 & ---      & ---      & 1.83E-08 & ---      & ---      \\
	72 & Ga & 2.82E-02 & ---      & ---      & 1.74E-08 & ---      & ---      \\
	65 & Ni & 4.65E-05 & ---      & ---      & 1.61E-08 & ---      & ---      \\
	71 & Ga & 3.45E-02 & ---      & ---      & 1.53E-08 & ---      & ---      \\
	55 & Mn & 3.50E-03 & 8.73E-03 & 4.26E-05 & 1.43E-08 & 7.70E-08 & 5.53E-10 \\
	69 & Zn & 1.06E-02 & ---      & ---      & 1.37E-08 & ---      & ---      \\
	58 & Fe & 3.27E-05 & ---      & ---      & 1.27E-08 & ---      & ---     \\
	\hline
	&     &             &           &   $\sum$$\dot{Y}^{ec}_e$ $\rightarrow$      &-1.01E-06 &-8.21E-06 & -4.89E-08
	\enddata
	
\end{deluxetable} \label{T1}

\begin{deluxetable}{cc|ccc|ccc}
	\tablenum{2}
	\tablecaption{Same as Table \ref{T1} but at conditions given below.}
	\tablewidth{0pt}
	\tablehead{
		\multicolumn1c{} & \multicolumn2c{$\rho$ = 1.45E+08,} &\multicolumn2c{$T_9$ = 3.8,}  &
		\multicolumn{2}{c}{$Y_e$ = 0.450} & \multicolumn1c{}  \\
		\hline
		\colhead{} & \colhead{} &\multicolumn3c{ {$\lambda_k^{ec}$}}  &
		\multicolumn{3}{c}{$\mid$$\dot{Y}^{ec}_{e(k)}$$\mid$}  \\
		\colhead{A} & \colhead{Symbol} & pn-QRPA & IPM & GTh &
		\colhead{pn-QRPA} & \colhead{IPM} & \colhead{GTh}
	}
	\startdata
	60 & Co & 1.04E-02 & 1.27E-02 & 2.76E-05 & 3.48E-08 & 1.27E-07 & 9.41E-10 \\
	59 & Co & 4.63E-04 & 6.57E-04 & 4.35E-08 & 2.41E-08 & 1.27E-07 & 7.60E-12 \\
	61 & Ni & 9.79E-04 & 1.20E-03 & 3.72E-07 & 9.23E-09 & 2.86E-08 & 2.20E-11 \\
	53 & Mn & 9.98E-03 & 8.97E-03 & ---      & 8.16E-09 & 1.37E-08 & ---      \\
	64 & Cu & 8.65E-02 & 3.69E-01 & ---      & 6.41E-09 & 4.25E-08 & ---      \\
	55 & Fe & 4.38E-03 & 6.00E-03 & 3.54E-04 & 5.68E-09 & 1.25E-08 & 1.23E-09 \\
	58 & Co & 1.69E-02 & 3.07E-02 & 4.00E-04 & 4.52E-09 & 1.48E-08 & 1.83E-09 \\
	56 & Fe & 2.24E-06 & 1.31E-06 & ---      & 3.74E-09 & 1.93E-09 & ---      \\
	57 & Fe & 2.33E-05 & 1.84E-05 & ---      & 3.32E-09 & 3.82E-09 & ---      \\
	63 & Cu & 1.00E-02 & 1.86E-02 & 9.89E-11 & 3.12E-09 & 1.53E-08 & 9.59E-11 \\
	55 & Mn & 1.26E-05 & 2.25E-05 & ---      & 2.56E-09 & 1.21E-08 & ---      \\
	53 & Cr & 8.47E-06 & 2.46E-06 & ---      & 1.77E-09 & 9.57E-10 & ---      \\
	51 & Cr & 3.09E-02 & 9.33E-03 & 8.47E-04 & 1.37E-09 & 1.36E-09 & 2.94E-10 \\
	54 & Mn & 4.06E-04 & 1.57E-02 & ---      & 1.33E-09 & 6.68E-08 & ---      \\
	65 & Cu & 5.75E-04 & ---      & ---      & 1.24E-09 & ---      & ---      \\
	57 & Co & 3.92E-03 & 1.29E-02 & 5.40E-04 & 7.59E-10 & 8.67E-09 & 4.97E-10 \\
	51 & V  & 1.53E-05 & 2.96E-05 & ---      & 7.02E-10 & 3.45E-09 & ---      \\
	49 & Ti & 4.40E-04 & ---      & ---      & 4.12E-10 & ---      & ---      \\
	52 & Cr & 3.80E-07 & ---      & ---      & 2.89E-10 & ---      & ---      \\
	60 & Ni & 4.15E-06 & 2.74E-05 & ---      & 2.67E-10 & 1.53E-09 & ---      \\
	49 & V  & 1.34E-01 & ---      & ---      & 2.54E-10 & ---      & ---      \\
	50 & V  & 5.35E-03 & 2.45E-02 & 1.36E-03 & 2.10E-10 & 2.60E-09 & 8.31E-10 \\
	56 & Mn & 3.08E-05 & 2.56E-04 & 3.36E-09 & 2.06E-10 & 3.22E-09 & 2.88E-13 \\
	59 & Ni & 7.55E-03 & ---      & 1.17E-03 & 1.83E-10 & ---      & 1.46E-10 \\
	62 & Cu & 3.02E-01 & 4.60E+00 & 1.07E-03 & 1.82E-10 & 3.77E-09 & 1.07E-11 \\
	62 & Ni & 3.62E-08 & ---      & ---      & 9.63E-11 & ---      & ---      \\
	61 & Cu & 5.66E-01 & ---      & 2.02E-03 & 7.55E-11 & ---      & 1.64E-12 \\
	48 & V  & 1.26E+02 & ---      & ---      & 7.28E-11 & ---      & ---      \\
	63 & Ni & 2.06E-06 & ---      & ---      & 5.73E-11 & ---      & ---      \\
	54 & Fe & 6.79E-04 & ---      & ---      & 4.76E-11 & ---      & ---     \\
	\hline
	&     &             &           &   $\sum$$\dot{Y}^{ec}_e$ $\rightarrow$      &-2.55E-08 &-4.96E-07 & -5.91E-09
	\enddata
	
\end{deluxetable} \label{T2}

\begin{deluxetable}{cc|ccc|ccc}
	\tablenum{3}
	\tablecaption{Same as Table \ref{T1} but at conditions given below. }
	\tablewidth{0pt}
	\tablehead{
		\multicolumn1c{} & \multicolumn2c{$\rho$ = 5.86E+07,} &\multicolumn2c{$T_9$ = 3.40,}  &
		\multicolumn{2}{c}{$Y_e$ = 0.470} & \multicolumn1c{}  \\
		\hline
		\colhead{} & \colhead{} &\multicolumn3c{ {$\lambda_k^{ec}$}}  &
		\multicolumn{3}{c}{$\mid$$\dot{Y}^{ec}_{e(k)}$$\mid$}  \\
		\colhead{A} & \colhead{Symbol} & pn-QRPA & IPM & GTh &
		\colhead{pn-QRPA} & \colhead{IPM} & \colhead{GTh}
	}
	\startdata
	58 & Ni & 6.10E-04 & 6.36E-04 & 7.42E-06 & 6.39E-07 & 5.97E-07 & 2.45E-08 \\
	55 & Fe & 1.09E-03 & 1.61E-03 & 5.01E-05 & 5.71E-07 & 1.52E-06 & 1.14E-07 \\
	53 & Mn & 3.27E-03 & 2.48E-03 & ---      & 3.95E-07 & 6.09E-07 & ---      \\
	55 & Co & 1.25E-01 & 1.41E-01 & 1.68E-03 & 3.89E-07 & 1.37E-06 & 1.73E-07 \\
	54 & Fe & 8.61E-05 & 3.11E-04 & ---      & 3.74E-07 & 1.21E-06 & ---      \\
	57 & Co & 1.94E-03 & 3.50E-03 & 9.35E-05 & 3.07E-07 & 2.18E-06 & 1.25E-07 \\
	57 & Ni & 5.47E-02 & 1.94E-02 & 5.94E-03 & 1.40E-07 & 5.84E-08 & 2.45E-07 \\
	56 & Co & 2.64E-02 & 7.40E-02 & 1.68E-03 & 1.23E-07 & 7.35E-07 & 1.89E-07 \\
	59 & Ni & 2.17E-03 & 4.37E-03 & 2.41E-04 & 8.81E-08 & 3.80E-07 & 8.69E-08 \\
	61 & Cu & 1.35E-01 & 3.93E-01 & 4.96E-04 & 4.00E-08 & 3.22E-07 & 1.83E-09 \\
	51 & Cr & 8.99E-03 & 2.81E-03 & 1.44E-04 & 1.71E-08 & 1.97E-08 & 3.61E-09 \\
	48 & V  & 3.47E+01 & ---      & ---      & 1.50E-08 & ---      & ---      \\
	56 & Ni & 2.01E-02 & 1.60E-02 & 1.98E-03 & 1.17E-08 & 7.76E-09 & 2.66E-08 \\
	58 & Co & 9.98E-03 & 1.04E-02 & 9.73E-05 & 1.09E-08 & 2.35E-08 & 2.60E-09 \\
	53 & Fe & 5.73E-02 & 2.04E-02 & ---      & 7.90E-09 & 9.70E-09 & ---      \\
	52 & Mn & 2.13E-02 & 2.85E-02 & ---      & 5.45E-09 & 1.62E-08 & ---      \\
	50 & Cr & 1.67E-04 & ---      & ---      & 3.04E-09 & ---      & ---      \\
	59 & Cu & 1.69E+00 & 1.01E+00 & ---      & 2.99E-09 & 4.62E-09 & ---      \\
	54 & Mn & 6.93E-04 & 5.13E-03 & ---      & 1.87E-09 & 2.05E-08 & ---      \\
	56 & Fe & 1.06E-07 & ---      & ---      & 9.98E-10 & ---      & ---      \\
	49 & V  & 3.51E-02 & ---      & ---      & 7.78E-10 & ---      & ---      \\
	62 & Cu & 6.14E-02 & 1.59E+00 & ---      & 4.70E-10 & 1.85E-08 & ---      \\
	60 & Cu & 2.02E-01 & 8.39E-01 & 4.07E-03 & 3.92E-10 & 4.30E-09 & 3.30E-10 \\
	60 & Ni & 2.30E-07 & 1.49E-06 & ---      & 3.61E-10 & 2.21E-09 & ---      \\
	51 & Mn & 1.43E-02 & ---      & ---      & 3.45E-10 & ---      & ---      \\
	62 & Zn & 2.05E-02 & ---      & ---      & 1.66E-10 & ---      & ---      \\
	52 & Fe & 1.50E-02 & ---      & ---      & 1.54E-10 & ---      & ---      \\
	49 & Cr & 3.67E-02 & ---      & ---      & 8.31E-11 & ---      & ---      \\
	61 & Ni & 8.20E-05 & ---      & ---      & 7.02E-11 & ---      & ---      \\
	64 & Zn & 7.76E-04 & ---      & ---      & 6.95E-11 & ---      & ---     \\
	\hline
	&     &             &           &   $\sum$$\dot{Y}^{ec}_e$ $\rightarrow$      &-3.15E-06 &-8.21E-06 & -9.96E-07
	\enddata
	
\end{deluxetable} \label{T3}

\begin{deluxetable}{cc|ccc|ccc}
	\tablenum{4}
	\tablecaption{Same as Table \ref{T1} but for \textit{bd} and at conditions given below. }
	\tablewidth{0pt}
	\tablehead{
		\multicolumn1c{} & \multicolumn2c{$\rho$ = 2.20E+09,} &\multicolumn2c{$T_9$ = 5.39,}  &
		\multicolumn{2}{c}{$Y_e$ = 0.425} & \multicolumn1c{}  \\
		\hline
		\colhead{} & \colhead{} &\multicolumn3c{ {$\lambda_k^{bd}$}}  &
		\multicolumn{3}{c}{$\mid$$\dot{Y}^{bd}_{e(k)}$$\mid$}  \\
		\colhead{A} & \colhead{Symbol} & pn-QRPA & IPM & GTh &
		\colhead{pn-QRPA} & \colhead{IPM} & \colhead{GTh}
	}
	\startdata
	65 & Co & 1.75E-01 & 8.17E-02 & ---      & 2.81E-06 & 4.40E-06 & ---      \\
	51 & Sc & 2.00E-01 & 3.44E-02 & 3.64E-03 & 1.38E-06 & 1.02E-06 & 8.66E-07 \\
	53 & Ti & 4.03E-02 & 1.47E-02 & 7.14E-04 & 1.24E-06 & 1.07E-06 & 8.66E-08 \\
	49 & Ca & 9.02E-03 & 9.31E-03 & 1.22E-03 & 1.07E-06 & 1.14E-06 & 3.91E-07 \\
	66 & Co & 1.43E+00 & 3.24E-01 & ---      & 9.24E-07 & 1.01E-06 & ---      \\
	62 & Fe & 2.66E-03 & 9.63E-03 & ---      & 7.21E-07 & 2.51E-06 & ---      \\
	59 & Mn & 2.02E-02 & 1.39E-01 & 2.06E-04 & 6.51E-07 & 7.93E-06 & 3.03E-08 \\
	67 & Ni & 6.53E-04 & 4.02E-03 & 7.88E-06 & 6.26E-07 & 2.85E-06 & 7.85E-09 \\
	64 & Co & 1.98E-02 & 1.14E-01 & ---      & 5.58E-07 & 2.56E-06 & ---      \\
	58 & Cr & 2.92E-02 & 1.64E-01 & ---      & 5.45E-07 & 5.03E-06 & ---      \\
	77 & Ga & 4.98E-02 & ---      & ---      & 5.21E-07 & ---      & ---      \\
	57 & Cr & 1.85E-02 & 1.20E-01 & ---      & 4.90E-07 & 9.17E-06 & ---      \\
	63 & Fe & 9.84E-02 & 3.54E-01 & 1.47E-02 & 4.36E-07 & 1.61E-06 & 4.27E-07 \\
	67 & Co & 3.24E+00 & ---      & ---      & 3.97E-07 & ---      & ---      \\
	50 & Sc & 6.41E-03 & 5.54E-02 & 4.19E-03 & 3.31E-07 & 5.16E-06 & 1.44E-05 \\
	60 & Mn & 5.71E-02 & 3.38E-01 & 7.78E-03 & 2.82E-07 & 1.22E-06 & 3.92E-08 \\
	51 & Ti & 5.02E-04 & ---      & ---      & 2.60E-07 & ---      & ---      \\
	71 & Cu & 4.33E-02 & ---      & ---      & 2.39E-07 & ---      & ---      \\
	61 & Fe & 1.63E-03 & 3.54E-02 & ---      & 1.70E-07 & 8.50E-06 & ---      \\
	68 & Ni & 1.93E-04 & ---      & ---      & 1.67E-07 & ---      & ---      \\
	54 & V  & 1.01E-02 & 6.08E-02 & 4.05E-03 & 1.56E-07 & 1.87E-06 & 2.72E-07 \\
	49 & Sc & 8.20E-04 & ---      & ---      & 1.55E-07 & ---      & ---      \\
	52 & Sc & 6.44E-01 & ---      & ---      & 1.27E-07 & ---      & ---      \\
	63 & Co & 1.27E-03 & 5.27E-03 & ---      & 1.27E-07 & 1.58E-06 & ---      \\
	75 & Ga & 2.82E-03 & ---      & ---      & 1.20E-07 & ---      & ---      \\
	72 & Cu & 3.94E-01 & ---      & ---      & 1.12E-07 & ---      & ---      \\
	55 & V  & 9.82E-03 & ---      & ---      & 1.11E-07 & ---      & ---      \\
	52 & Ti & 1.70E-04 & 4.49E-04 & ---      & 1.10E-07 & 2.69E-07 & ---      \\
	56 & V  & 4.61E-01 & ---      & ---      & 1.09E-07 & ---      & ---      \\
	69 & Ni & 6.95E-03 & ---      & ---      & 1.08E-07 & ---      & ---     \\
	\hline
	&     &             &           &   $\sum$$\dot{Y}^{bd}_e$ $\rightarrow$      &1.59E-05 &6.87E-5 & 3.16E-06
	\enddata
	
\end{deluxetable} \label{T4}

\begin{deluxetable}{cc|ccc|ccc}
	\tablenum{5}
	\tablecaption{Same as Table \ref{T1} but for \textit{bd} and at conditions given below.}
	\tablewidth{0pt}
	\tablehead{
		\multicolumn1c{} & \multicolumn2c{$\rho$ = 3.30E+08,} &\multicolumn2c{$T_9$ = 4.24,}  &
		\multicolumn{2}{c}{$Y_e$ = 0.440} & \multicolumn1c{}  \\
		\hline
		\colhead{} & \colhead{} &\multicolumn3c{ {$\lambda_k^{bd}$}}  &
		\multicolumn{3}{c}{$\mid$$\dot{Y}^{bd}_{e(k)}$$\mid$}  \\
		\colhead{A} & \colhead{Symbol} & pn-QRPA & IPM & GTh &
		\colhead{pn-QRPA} & \colhead{IPM} & \colhead{GTh}
	}
	\startdata
	67 & Ni & 2.62E-02 & 1.20E-02 & 5.29E-04 & 8.52E-06 & 1.05E-07 & 5.70E-09 \\
	49 & Sc & 3.55E-02 & ---      & ---      & 2.00E-06 & ---      & ---      \\
	63 & Co & 3.16E-02 & 1.41E-02 & 3.52E-04 & 1.41E-06 & 7.46E-07 & 2.00E-08 \\
	50 & Sc & 3.08E-01 & 1.79E-01 & 4.10E-02 & 1.21E-06 & 2.08E-07 & 2.04E-07 \\
	65 & Co & 7.46E-01 & 1.39E-01 & ---      & 1.06E-06 & 2.02E-08 & ---      \\
	59 & Mn & 1.44E-01 & 1.58E-01 & 5.40E-03 & 8.79E-07 & 2.89E-07 & 3.20E-08 \\
	64 & Co & 2.68E-01 & 2.05E-01 & ---      & 8.62E-07 & 5.42E-08 & ---      \\
	58 & Cr & 4.04E-01 & 2.79E-01 & ---      & 7.47E-07 & 2.88E-08 & ---      \\
	69 & Cu & 3.53E-02 & ---      & ---      & 5.90E-07 & ---      & ---      \\
	61 & Fe & 1.57E-02 & 6.44E-02 & 7.45E-04 & 5.29E-07 & 1.36E-06 & 3.45E-08 \\
	51 & Ti & 1.76E-03 & 7.13E-04 & ---      & 4.32E-07 & 1.40E-07 & ---      \\
	57 & Cr & 1.01E-01 & 1.49E-01 & ---      & 3.66E-07 & 2.00E-07 & ---      \\
	62 & Fe & 2.75E-03 & 4.72E-02 & ---      & 2.70E-07 & 2.90E-07 & ---      \\
	51 & Sc & 8.65E-01 & 1.32E-01 & 4.73E-02 & 2.37E-07 & 6.16E-09 & 8.17E-07 \\
	55 & Cr & 2.79E-03 & 1.06E-03 & 1.12E-05 & 2.19E-07 & 2.49E-07 & 5.57E-09 \\
	71 & Cu & 7.33E-01 & ---      & ---      & 1.45E-07 & ---      & ---      \\
	63 & Fe & 5.82E-01 & ---      & 1.68E-01 & 1.12E-07 & ---      & 5.33E-09 \\
	53 & Ti & 3.92E-02 & 3.19E-02 & 1.95E-02 & 1.02E-07 & 1.04E-08 & 1.24E-08 \\
	49 & Ca & 1.39E-02 & 4.18E-02 & 2.84E-02 & 9.42E-08 & 7.27E-09 & 1.35E-08 \\
	54 & V  & 5.38E-02 & 1.90E-01 & 3.85E-02 & 8.67E-08 & 2.04E-07 & 1.15E-07 \\
	75 & Ga & 3.21E-02 & ---      & ---      & 8.41E-08 & ---      & ---      \\
	60 & Mn & 3.82E-01 & ---      & ---      & 7.83E-08 & ---      & ---      \\
	70 & Cu & 3.34E-01 & ---      & ---      & 7.73E-08 & ---      & ---      \\
	57 & Mn & 2.37E-03 & 5.32E-04 & ---      & 7.68E-08 & 8.02E-08 & ---      \\
	58 & Mn & 1.14E-02 & 1.39E-01 & 5.80E-08 & 7.32E-08 & 6.30E-07 & 8.70E-08 \\
	53 & V  & 2.26E-03 & 2.98E-03 & 8.82E-05 & 6.64E-08 & 2.63E-07 & 2.05E-08 \\
	55 & V  & 5.83E-02 & 8.52E-02 & ---      & 5.19E-08 & 1.97E-08 & ---      \\
	54 & Ti & 1.33E-01 & ---      & ---      & 4.41E-08 & ---      & ---      \\
	65 & Ni & 1.71E-04 & 2.49E-03 & ---      & 4.22E-08 & 5.36E-07 & ---      \\
	56 & Cr & 8.32E-05 & 1.14E-04 & ---      & 4.17E-08 & 2.01E-08 & ---     \\
	\hline
	&     &             &           &   $\sum$$\dot{Y}^{bd}_e$ $\rightarrow$      &2.08E-06 &1.01E-05 & 6.39E-07
	\enddata
	
\end{deluxetable} \label{T5}

\begin{deluxetable}{cc|ccc|ccc}
	\tablenum{6}
	\tablecaption{Same as Table \ref{T1} but for \textit{bd} and at conditions given below. }
	\tablewidth{0pt}
	\tablehead{
		\multicolumn1c{} & \multicolumn2c{$\rho$ = 3.86E+07,} &\multicolumn2c{$T_9$ = 3.40,}  &
		\multicolumn{2}{c}{$Y_e$ = 0.470} & \multicolumn1c{}  \\
		\hline
		\colhead{} & \colhead{} &\multicolumn3c{ {$\lambda_k^{bd}$}}  &
		\multicolumn{3}{c}{$\mid$$\dot{Y}^{bd}_{e(k)}$$\mid$}  \\
		\colhead{A} & \colhead{Symbol} & pn-QRPA & IPM & GTh &
		\colhead{pn-QRPA} & \colhead{IPM} & \colhead{GTh}
	}
	\startdata
	55 & Mn & 1.08E-07 & 3.68E-07 & ---      & 1.78E-13 & 1.75E-12 & ---      \\
	57 & Fe & 3.34E-08 & 1.10E-05 & ---      & 9.45E-14 & 5.04E-11 & ---      \\
	58 & Fe & 2.90E-08 & 1.09E-07 & ---      & 4.14E-14 & 1.55E-13 & ---      \\
	52 & V  & 8.71E-03 & 1.60E-02 & 4.32E-04 & 3.93E-14 & 1.33E-13 & 2.76E-14 \\
	53 & Cr & 3.75E-08 & 5.93E-06 & ---      & 2.55E-14 & 8.22E-12 & ---      \\
	54 & Cr & 2.02E-07 & 2.90E-07 & ---      & 1.70E-14 & 2.46E-14 & ---      \\
	59 & Co & 3.86E-09 & 8.11E-08 & ---      & 8.90E-15 & 7.69E-13 & ---      \\
	61 & Co & 2.86E-04 & 1.36E-03 & ---      & 5.01E-15 & 1.02E-13 & ---      \\
	57 & Mn & 7.43E-03 & ---      & 2.97E-05 & 4.24E-15 & ---      & 1.72E-16 \\
	56 & Fe & 4.32E-13 & 1.19E-10 & ---      & 4.08E-15 & 1.07E-12 & ---      \\
	54 & Mn & 1.12E-09 & 8.81E-06 & ---      & 3.01E-15 & 3.52E-11 & ---      \\
	53 & Mn & 1.76E-11 & ---      & ---      & 2.12E-15 & ---      & ---      \\
	56 & Mn & 7.41E-06 & 1.09E-02 & 2.37E-04 & 1.24E-15 & 3.87E-12 & 4.23E-13 \\
	55 & Cr & 4.84E-03 & ---      & ---      & 1.04E-15 & ---      & ---      \\
	60 & Co & 2.03E-06 & 4.66E-03 & 4.93E-05 & 1.03E-15 & 7.94E-12 & 2.17E-13 \\
	52 & Cr & 1.48E-12 & ---      & ---      & 9.19E-16 & ---      & ---      \\
	59 & Fe & 4.28E-05 & 6.95E-03 & 7.29E-07 & 6.66E-16 & 2.43E-13 & 1.15E-17 \\
	57 & Co & 2.27E-12 & 1.34E-09 & ---      & 3.59E-16 & 8.36E-13 & ---      \\
	51 & V  & 5.96E-09 & ---      & ---      & 2.87E-16 & ---      & ---      \\
	55 & Fe & 5.01E-13 & 1.30E-10 & ---      & 2.61E-16 & 1.23E-13 & ---      \\
	51 & Cr & 1.05E-10 & ---      & ---      & 1.99E-16 & ---      & ---      \\
	53 & V  & 6.95E-03 & ---      & ---      & 5.39E-17 & ---      & ---      \\
	49 & Ti & 1.96E-07 & ---      & ---      & 4.94E-17 & ---      & ---      \\
	61 & Ni & 5.09E-11 & ---      & ---      & 4.36E-17 & ---      & ---      \\
	51 & Ti & 2.89E-03 & ---      & ---      & 3.68E-17 & ---      & ---      \\
	63 & Ni & 4.50E-07 & 3.16E-04 & ---      & 3.53E-17 & 2.90E-14 & ---      \\
	58 & Co & 2.96E-11 & 5.67E-06 & ---      & 3.22E-17 & 1.27E-11 & ---      \\
	50 & V  & 7.87E-09 & 1.89E-05 & 3.2E-09  & 2.38E-17 & 1.79E-13 & 2.06E-16 \\
	49 & Sc & 8.30E-02 & ---      & ---      & 2.31E-17 & ---      & ---      \\
	62 & Ni & 7.36E-12 & ---      & ---      & 2.07E-17 & ---      & ---     \\
	\hline
	&     &             &           &   $\sum$$\dot{Y}^{bd}_e$ $\rightarrow$      &4.30E-13 &1.24E-10 & 6.68E-13
	\enddata
	
\end{deluxetable} \label{T6}


\begin{deluxetable}{ccc|ccc||ccc|ccc}
	\tablenum{7}
	\tablecaption{List of top 50 most important \textit{ec} and \textit{bd} nuclei averaged throughout the stellar trajectory for $0.500 > {Y_e} > 0.400$. The ranking paramter $\mathring{R}_{p}$ is defined in text. Nuclei marked with asterisk are new entries not to be seen in the list compiled by IPM.  }
	\tablewidth{0pt}
	\tablehead{
		\multicolumn{6}{c}{\textit{ec} nuclei}  &
		\multicolumn{6}{c}{\textit{bd} nulcei} \\
		A & Symbol & $\mathring{R}_{p}$ &A & Symbol & $\mathring{R}_{p}$ &A & Symbol & $\mathring{R}_{p}$ &A & Symbol & $\mathring{R}_{p}$
	}
	\startdata
	56 & Mn & 1.11E+01 & 69 & Cu & 3.69E-02 & 67 & Ni & 5.34E-01 & 63 & Fe$^\ast$ & 1.39E-02 \\
	52 & V  & 6.32E-01 & 54 & Cr & 3.63E-02 & 49 & Sc$^\ast$ & 1.33E-01 & 66 & Co & 1.35E-02 \\
	67 & Cu & 5.57E-01 & 81 & Ge$^\ast$ & 3.39E-02 & 65 & Co & 1.03E-01 & 58 & Fe & 1.24E-02 \\
	60 & Co & 4.03E-01 & 78 & Ga & 3.37E-02 & 63 & Co & 9.17E-02 & 71 & Cu$^\ast$ & 1.22E-02 \\
	53 & Mn & 2.51E-01 & 64 & Cu & 3.22E-02 & 50 & Sc & 8.56E-02 & 50 & Ca & 1.11E-02 \\
	49 & Sc & 2.40E-01 & 57 & Co & 3.19E-02 & 59 & Mn & 6.70E-02 & 60 & Mn & 8.93E-03 \\
	66 & Cu & 2.09E-01 & 63 & Cu & 3.04E-02 & 53 & Mn$^\ast$ & 6.56E-02 & 77 & Ga$^\ast$ & 8.82E-03 \\
	50 & Sc & 1.75E-01 & 58 & Ni & 3.02E-02 & 64 & Co & 6.50E-02 & 54 & V  & 8.20E-03 \\
	55 & Fe & 1.68E-01 & 57 & Fe & 2.96E-02 & 49 & Ca$^\ast$ & 6.39E-02 & 53 & Cr & 7.87E-03 \\
	59 & Co & 1.53E-01 & 77 & Ge & 2.88E-02 & 55 & Mn & 5.64E-02 & 75 & Ga$^\ast$ & 6.91E-03 \\
	79 & Ge & 1.11E-01 & 64 & Ni & 2.79E-02 & 58 & Cr$^\ast$ & 5.47E-02 & 79 & Ga$^\ast$ & 6.23E-03 \\
	77 & Ga & 7.89E-02 & 58 & Co & 2.76E-02 & 69 & Cu$^\ast$ & 3.73E-02 & 70 & Cu$^\ast$ & 6.06E-03 \\
	61 & Ni & 7.60E-02 & 83 & Se & 2.73E-02 & 61 & Fe & 3.66E-02 & 54 & Cr & 5.92E-03 \\
	78 & Ge$^\ast$ & 7.46E-02 & 53 & Cr & 2.62E-02 & 51 & Ti & 3.54E-02 & 56 & Fe & 5.87E-03 \\
	83 & As & 6.85E-02 & 48 & Sc$^\ast$ & 2.57E-02 & 51 & Sc & 3.27E-02 & 53 & V  & 5.84E-03 \\
	51 & Ti & 6.78E-02 & 51 & Cr & 2.51E-02 & 57 & Fe & 3.12E-02 & 58 & Mn & 5.11E-03 \\
	67 & Ni & 6.67E-02 & 65 & Ni & 2.48E-02 & 57 & Cr & 3.08E-02 & 55 & V  & 4.90E-03 \\
	56 & Ni & 6.17E-02 & 55 & Cr & 2.40E-02 & 55 & Cr & 2.79E-02 & 51 & Ca & 4.88E-03 \\
	68 & Cu & 4.62E-02 & 64 & Co & 2.31E-02 & 62 & Fe & 2.66E-02 & 54 & Ti & 4.66E-03 \\
	82 & As & 4.60E-02 & 57 & Mn & 2.26E-02 & 57 & Mn & 2.42E-02 & 61 & Co & 4.63E-03 \\
	56 & Fe & 4.55E-02 & 71 & Cu & 2.23E-02 & 57 & Co$^\ast$ & 2.14E-02 & 51 & Cr$^\ast$ & 4.25E-03 \\
	65 & Cu & 4.48E-02 & 55 & Mn & 2.07E-02 & 53 & Ti & 2.12E-02 & 68 & Ni & 3.96E-03 \\
	55 & Co & 4.03E-02 & 53 & V  & 2.05E-02 & 52 & V  & 1.80E-02 & 56 & Cr & 3.70E-03 \\
	75 & Ga$^\ast$ & 3.83E-02 & 73 & Ga$^\ast$ & 2.03E-02 & 55 & Fe$^\ast$ & 1.78E-02 & 68 & Co & 3.65E-03 \\
	57 & Ni & 3.69E-02 & 51 & V  & 1.96E-02 & 67 & Co$^\ast$ & 1.44E-02 & 54 & Mn & 3.37E-03
	\enddata
	
\end{deluxetable} \label{T7}


\newpage
\clearpage
\bibliography{ApJ}{}
\bibliographystyle{aasjournal}
\clearpage

\end{document}